\documentclass[11pt,letterpaper,final]{article}
\usepackage{emnlp2016}
\usepackage{times}
\usepackage{latexsym}
\usepackage{comment}
\usepackage{amsmath}
\usepackage{url}
\usepackage{graphicx}
\usepackage{color}
\usepackage{multirow}
\usepackage{hhline}
\usepackage{verbatim}
\usepackage{bm}
\usepackage{multirow}
\graphicspath{ {images/} }

\emnlpfinalcopy

\def\emnlppaperid{675}

\title{TweeTime: A Minimally Supervised Method for Recognizing and Normalizing Time Expressions in Twitter}

\author{Jeniya Tabassum, Alan Ritter and Wei Xu\\
	    Computer Science and Engineering\\
	    Ohio State University\\
	    {\tt \{tabassum.13,ritter.1492,xu.1265\}@osu.edu}
  }

\date{}

\begin{document}

\maketitle

\begin{abstract}
We describe TweeTIME, a temporal tagger for recognizing and normalizing time expressions in Twitter.  Most previous work in social media analysis has to rely on temporal resolvers that are designed for well-edited text, and therefore suffer from reduced performance due to domain mismatch. We present a minimally supervised method that learns from large quantities of unlabeled data and requires no hand-engineered rules or hand-annotated training corpora.  TweeTIME achieves 0.68 F1 score on the end-to-end task of resolving date expressions, outperforming a broad range of state-of-the-art systems.\footnote{Our code and data are publicly available at \url{https://github.com/jeniyat/TweeTime}.}


\end{abstract}

\section{Introduction}

Temporal expressions are words or phrases that refer to dates, times or durations.  Resolving time expressions is an important task in information extraction (IE) that enables downstream applications such as calendars or timelines of events \cite{derczynski2013temporal,Do2012,ritter2012open,ling2010temporal}, knowledge base population \cite{ji2011overview}, information retrieval \cite{alonso2007value}, automatically scheduling meetings from email and more. Previous work in this area has applied rule-based systems \cite{mani2000robust,bethard:2013:EMNLP,chambers2013tempeval} or supervised machine learning on small collections of hand-annotated news documents \cite{angeli2012parsing,lee2014context}. 

\begin{figure}
    \centering
    \includegraphics[width=0.38\textwidth]{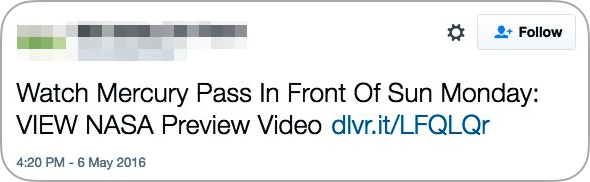}
    \caption{A tweet published on {\em Friday 5/6/2016} that contains the temporal expression {\em Monday} referring to the date of the event ({\em 5/9/2016}), which a generic temporal tagger failed to resolve correctly.}
    \label{fig:exampletweet1}
\end{figure}

\begin{figure}
    \centering
    \includegraphics[width=0.4\textwidth]{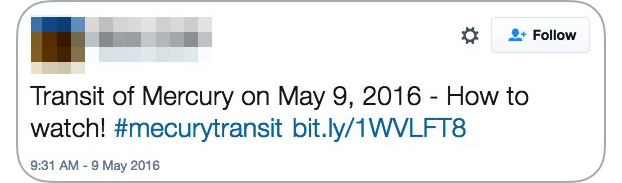}
    \caption{A tweet that contains a simple explicit time mention and an event ({\em Mercury, 5/9/2016}) that can be identified by an open-domain information extraction system.}
    \label{fig:exampletweet2}
\end{figure}

Social media especially contains time-sensitive information and requires accurate temporal analysis, for example, for detecting real-time cybersecurity events \cite{ritter2015weakly,changexpectation}, disease outbreaks \cite{kanhabua2012supporting} and extracting personal information \cite{schwartz2015extracting}. However, most work on social media simply uses generic temporal resolvers and therefore suffers from suboptimal performance. Recent work on temporal resolution focuses primarily on news articles and clinical texts \cite{uzzaman2013tempeval,bethard2016tempeval}. 

Resolving time expressions in social media is a non-trivial problem. Besides many spelling variations, time expressions are more likely to refer to future dates than in newswire.  For the example in Figure \ref{fig:exampletweet1}, we need to recognize that {\em Monday} refers to the upcoming Monday and not the previous one to resolve to its correct normalized date ({\em5/9/2016}). We also need to identify that the word {\em Sun} is not referring to a Sunday in this context.

\begin{figure*}[!ht]
    \centering
    \includegraphics[width=\textwidth]{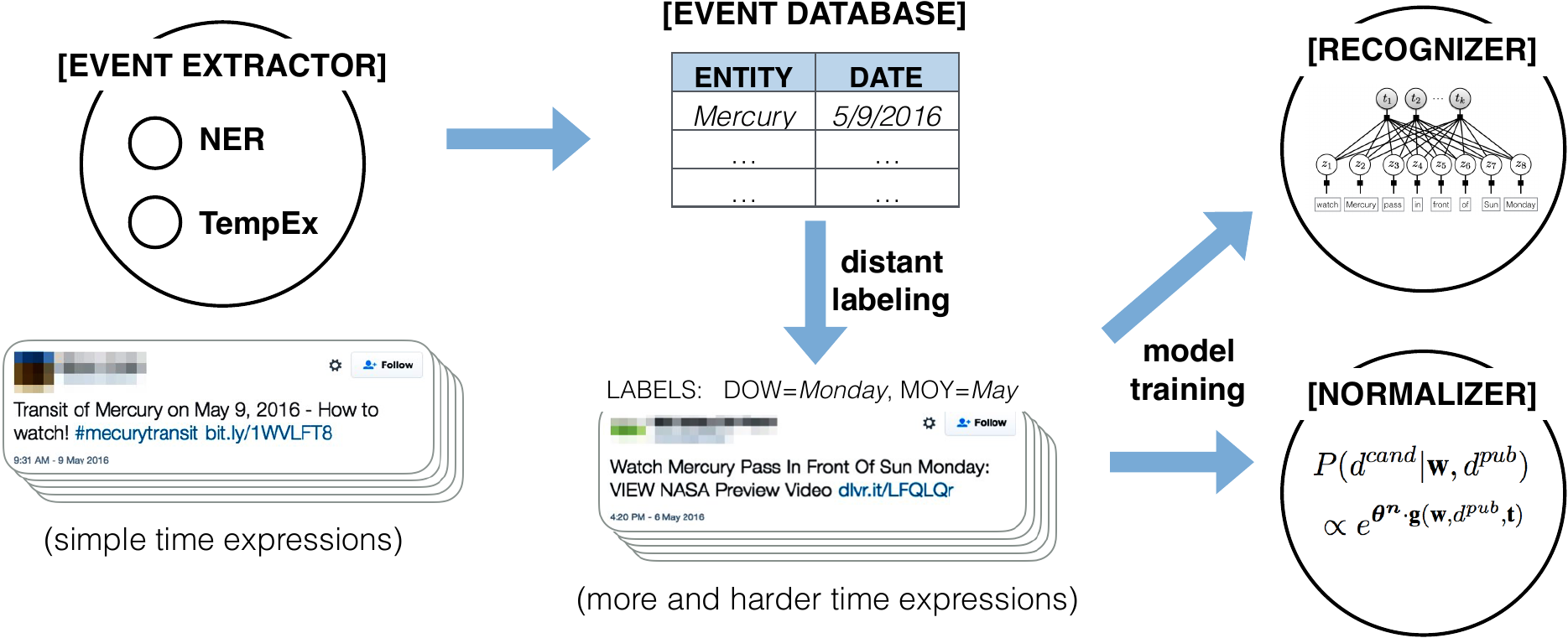}
    \caption{TweeTIME system diagram of model training.}
    \label{fig:systemdiagram}
\end{figure*}

In this paper, we present a new minimally supervised approach to temporal resolution that requires no in-domain annotation or hand-crafted rules, instead learning from large quantities of unlabeled text in conjunction with a database of known events.  Our approach is capable of learning robust time expression models adapted to the informal style of text found on social media. 


For popular events, some related tweets (e.g. Figure \ref{fig:exampletweet2}) may contain explicit or other simple time mentions that can be captured by a generic temporal tagger. An open-domain information extraction system \cite{ritter2012open} can then identify events (e.g. [{\it Mercury}, {\it 5/9/2016}]) by aggregating those tweets. To automatically generate temporally annotated data for training, we make the following novel {\it distant supervision assumption}:\footnote{We focus on resolving dates, arguably the most important and frequent category of time expressions in social media data, and leave other phenomenon such as times and durations to traditional methods or future work.}
\begin{quotation}
\noindent
Tweets posted near the time of a known event that mention central entities are likely to contain time expressions that refer to the date of the event.
\end{quotation}
Based on this assumption, tweets that contain the same named entity (e.g. Figure \ref{fig:exampletweet1}) are heuristically labeled as training data.  Each tweet is associated with multiple overlapping labels that indicate the day of the week, day of the month, whether the event is in the past or future and other time properties of the event date in relation to the tweet's creation date.  In order to learn a tagger that can recognize temporal expressions at the word-level, we present a multiple-instance learning approach to model sentence and word-level tags jointly and handle overlapping labels. Using heuristically labeled data and the temporal tags predicted by the multiple-instance learning model as input, we then train a log-linear model that normalizes time expressions to calendar dates.

Building on top of the multiple-instance learning model, we further improve performance using a missing data model that addresses the problem of errors introduced during the heuristic labeling process. Our best model achieves a 0.68 F1 score when resolving date mentions in Twitter. This is a 17\% increase over SUTime \cite{chang2012sutime}, outperforming other state-of-the-art time expression resolvers HeidelTime \cite{strotgen2013multilingual}, TempEX \cite{mani2000robust} and UWTime \cite{lee2014context} as well. Our approach also produces a confidence score that allows us to trade recall for precision. To the best of our knowledge, TweeTIME is the first time resolver designed specifically for social media data.\footnote{The closest work is HeidelTime's colloquial English version \cite{strotgen2012temporal} developed from annotated SMS data and slang dictionary. Our TweeTIME significantly outperforms on Twitter data.} This is also the first time that distant supervision is successfully applied for end-to-end temporal recognition and normalization. Previous distant supervision approaches \cite{angeli2012parsing,angeli2013language} only address the normalization problem, assuming gold time mentions are available at test time.


\section{System Overview}
\label{extract_resolve}

Our TweeTIME system consists of two major components as shown in Figure \ref{fig:systemdiagram}:

\begin{enumerate}
\item A {\bf Temporal Recognizer} which identifies time expressions (e.g. {\em Monday}) in English text and outputs 5 different temporal types (described in Table \ref{tab:tag_desc}) indicating timeline direction, month of year, date of month, day of week or no temporal information (NA). It is realized as a multiple-instance learning model, and in an enhanced version, as a missing data model. 
\item A {\bf Temporal Normalizer} that takes a tweet with its creation time and temporal expressions tagged by the above step as input, and outputs their normalized forms (e.g. {\em Monday} $\rightarrow$ {\em 5/9/2016}). It is a log-linear model that uses both lexical features and temporal tags.
\end{enumerate}

\begin{table}[!ht]
\centering
\begin{tabular}{c|c} 
 \hline
  {\bf Temporal Types} & {\bf Possible Values (tags)} \\
 \hline
 \hline
 Timeline (TL) & $past, present, future$ \\ 
 \hline
  Day of Week (DOW) & $Mon, Tue, \ldots, Sun$ \\  
 \hline
  Day of Month (DOM) & $1, 2, 3, \ldots , 31$ \\  
 \hline
  Month of Year (MOY) & $Jan, Feb, \ldots, Dec$ \\  
 \hline
  None (NA) & $NA$ \\  
 \hline
\end{tabular}
\caption{Our Temporal Recognizer can extract five different temporal types and assign one of their values to each word of a tweet.}
\label{tab:tag_desc}
\end{table}

To train these two models without corpora manually annotated with time expressions, we leverage a large database of known events as distant supervision. The event database is extracted automatically from Twitter using the open-domain IE system proposed by Ritter et al. \shortcite{ritter2012open}. Each event consists of one or more named entities, in addition to the date on which the event takes place, for example [{\em Mercury}, {\em 5/9/2016}].  Tweets are first processed by a Twitter named entity recognizer \cite{ritter2011named}, and a generic date resolver \cite{mani2000robust}. Events are then extracted based on the strength of association between each named entity and calendar date, as measured by a $G^2$ test on their co-occurrence counts.  More details of the {\bf Event Extractor} can be found in Section \ref{sec:event_extraction}. 

The following two sections describe the details of our {\bf Temporal Recognizer} and {\bf Temporal Normalizer} separately. 

\section{Distant Supervision for Recognizing Time Expressions}
\label{sec:tagging}
The goal of the recognizer is to predict the temporal tag of each word, given a sentence (or a tweet) $\textbf{w}=w_1,\ldots,w_n$. We propose a multiple-instance learning model and a missing data model that are capable of learning word-level taggers given only sentence-level labels.

Our recognizer module in  is built using a database of known events as {\it distant supervision}. We assume tweets published around the time of a known event that mention a central entity are also likely to contain time expressions referring to the event's date. For each event, such as [{\it Mercury}, {\it 5/9/2016}], we gather all tweets that contain the central entity {\it Mercury} and are posted within 7 days of {\it 5/9/2016}. We then label each tweet based on the event date in addition to the tweet's creation date.
The sentence-level temporal tags for the tweet in Figure \ref{fig:exampletweet1} are: TL=$future$, DOW=$Mon$, DOM=$9$, MOY=$May$. 


\subsection{Multiple-Instance Learning Temporal Tagging Model (MultiT)}
Unlike supervised learning, where labeled instances are provided to the learner, in multiple instance learning scenarios \cite{dietterich1997solving}, the learner is only provided with bags of instances labeled as either positive (where at least one instance is positive) or all negative.  This is a close match to our problem setting, in which sentences are labeled with tags that should be assigned to one or more words.

\begin{figure}
    \centering
    \includegraphics[width=0.5\textwidth]{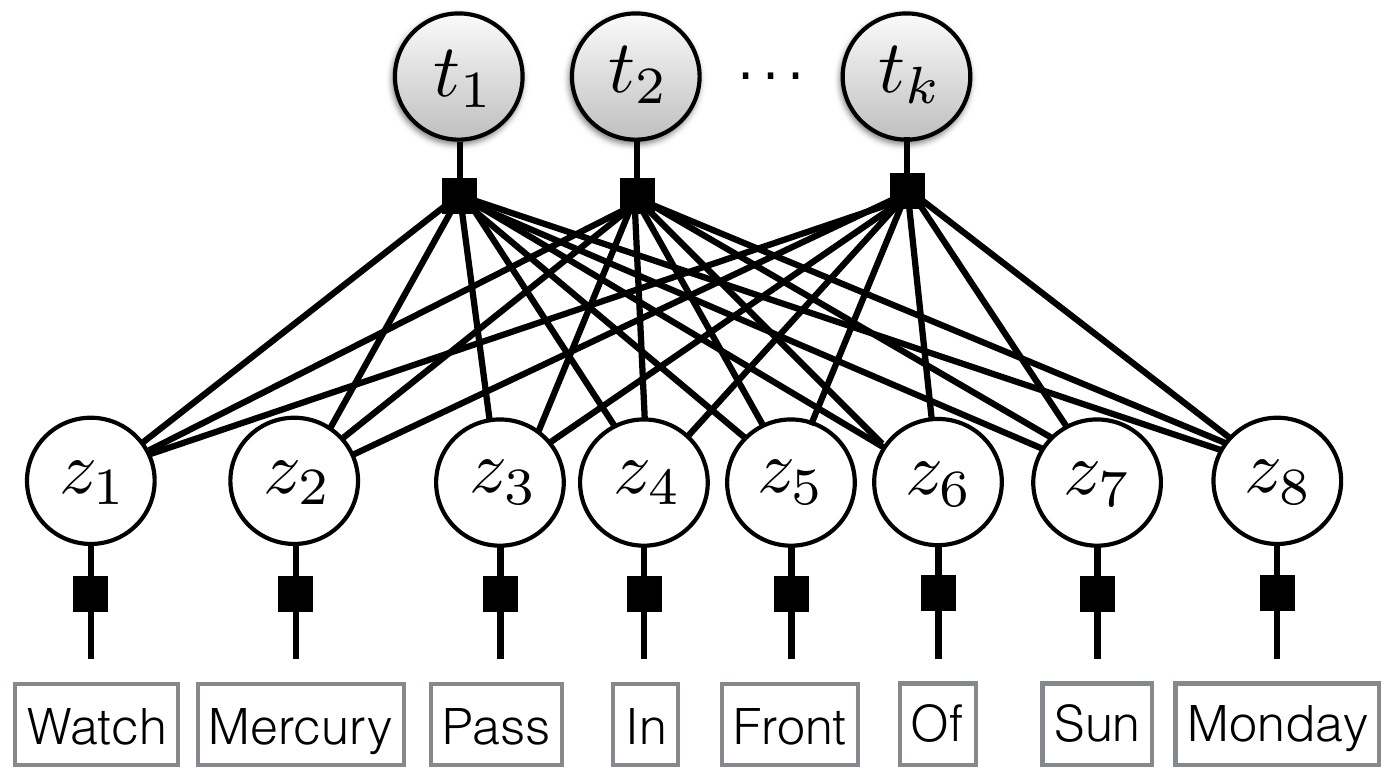}
    \caption{Multiple-Instance Learning Temporal Tagging Model -- our approach to learn a word-level tagging model given only sentence-level labels.  In this example a sentence-level variable $t_a=1$ indicates the temporal tag DOW=$Mon$ must be present and $t_b=1$ indicates that the target date is in the future (TL=$future$). The multiple instance learning assumption implies that at least one word must be tagged with each of these present temporal tags. For example, ideally after training, the model will learn to assign $z_8$ to tag $a$ and $z_1$ to tag $b$.}
    \label{fig:factor_graph}
\end{figure}

We represent sentences and their labels using a graphical model that is  divided into word-level and sentence-level variables (as shown in Figure \ref{fig:factor_graph}). Unlike the standard supervised tagging problem, we never directly observe the words' tags ($\textbf{z} = z_1, \ldots , z_n$) during learning.  Instead, they are latent and we only observe the date of an event mentioned in the text, from which we derive sentence-level binary variables $\textbf{t} = t_1, \ldots , t_k$ corresponding to temporal tags for the sentence. Following previous work on multiple-instance learning \cite{Hoffmann_multir,Xu-EtAl-2014:TACL}, we model the connection between sentence-level labels and word-level tags using a set of deterministic-OR factors $\phi^{sent}$.

The overall conditional probability of our model is defined as:
\begin{equation}
\begin{split}
& P(\textbf{t}, \textbf{z} | \textbf{w}; \bm{\theta^r})\\ 
& = \frac{1}{Z} \prod_{i=1}^k \phi^{sent}(t_i, \textbf{z}) \times \prod_{j=1}^n \phi^{word}(z_j,w_j) \\
& = \frac{1}{Z} \prod_{i=1}^k \phi^{sent}(t_i, \textbf{z}) \times \prod_{j=1}^n e^{\bm{\theta^r} \cdot \textbf{f}(z_j,w_j)}
\end{split}
\end{equation}
where $\textbf{f}(z_j,w_j)$ is a feature vector and
\begin{equation}
\phi^{sent}(t_i, \textbf{z})  = \begin{cases}
 1 & \text{if }t_i=true \land \exists j: z_j=i\\
 1 & \text{if }t_i=false \land \forall j: z_j \ne i\\
 0 & \text{otherwise}
\end{cases}
\end{equation}

We include a standard set of tagging features that includes word shape and identity in addition to prefixes and suffixes.  To learn parameters $\bm{\theta^r}$ of the Temporal Tagger, we maximize the likelihood of the sentence-level heuristic labels conditioned on observed words over all tweets in the training corpus. Given a training instance $\textbf{w}$ with label $\textbf{t}$, the gradient of the conditional log-likelihood with respect to the parameters is:
\begin{equation}
\begin{split}
\nabla P(\textbf{t} | & \textbf{w}) = \sum_{\textbf{z}} P(\textbf{z} | \textbf{w}, \textbf{t}; \bm{\theta^r}) \cdot \textbf{f}(\textbf{z}, \textbf{w}) \\
& - \sum_{\textbf{t}, \textbf{z}} P(\textbf{t}, \textbf{z} | \textbf{w}; \bm{\theta^r}) \cdot \textbf{f}(\textbf{z}, \textbf{w}) 
\end{split}
\end{equation}
This gradient is the difference of two conditional expectations over the feature vector $\textbf{f}$: a ``clamped'' expectation that is conditioned on the observed words and tags ($\textbf{w}$, $\textbf{t}$) and a ``free'' expectation that is only conditioned on the words in the text, $\textbf{w}$, and ignores the sentence-level labels. To make the inference tractable, we use a Viterbi approximation that replaces the expectations with maximization.  Because each sentence corresponds to more than one temporal tag, the maximization of the ``clamped'' maximization is somewhat challenging to compute. We use the approximate inference algorithm of Hoffmann et al. \shortcite{Hoffmann_multir}, that views inference as a weighted set cover problem, with worst case running time $(|T| \cdot |W|)$, where $|T|$ is the number of all possible temporal tag values and $|W|$ is the number of words in a sentence.

 \subsection{Missing Data Temporal Tagging Model (MiDaT)}\label{sec:missing_data}

While the multiple-instance learning assumption works well much of the time, it can easily be violated -- there are many tweets that mention entities involved in an event but that never explicitly mention its date.

The missing data modeling approach to weakly supervised learning proposed by Ritter et. al. \shortcite{ritter13} addresses this problem by relaxing the hard constraints of deterministic-OR factors, such as those described above, as soft constraints.  Our missing-data model for weakly supervised tagging splits the sentence-level variables, $t$ into two parts : $m$ which represents whether a temporal tag is mentioned by at least one word of the tweet, and $t'$ which represents whether a temporal tag can be derived from the event date.  A set of pairwise potentials $\psi(m_j, t'_j)$ are introduced that encourage (but don't strictly require) agreement between $m_j$ and $t'_j$, that is:
\begin{equation}
    \psi(m_j, t'_j)=
    \begin{cases}
      \alpha_{p}, \text{ if }\ t'_j \ne	 m_j \\
      \alpha_{r}, \text{ if }\ t'_j = m_j \\
    \end{cases}
\end{equation}

Here, $\alpha_{p}$ (Penalty), and $\alpha_{r}$ (Reward) are parameters for the MiDaT model. $\alpha_{p}$ is the penalty for extracting a temporal tag that is not related to the event-date and  $\alpha_{r}$ is the reward for extracting a tag that matches the date.  

During learning, if the local classifier is very confident, it is possible for a word to be labeled with a tag that is not derived from the event-date, and also for a sentence-level tag to be ignored, although either case will be penalized by the agreement potentials, $\psi(m_j, t'_j)$, in the global objective.  We use a local-search approach to inference that was empirically demonstrated to nearly always yield exact solutions by Ritter et. al. \shortcite{ritter13}.

\section{A Log-Linear Model for Normalizing Time Expressions}
The Temporal Normalizer is built using a log-linear model which takes the tags $\textbf{t}$ produced by the Temporal Recognizer as input and outputs one or more dates mentioned in a tweet. We formulate date normalization as a binary classification problem: given a tweet $\textbf{w}$ published on date $d^{pub}$, we consider 22 candidate target dates ($\textbf{w}$, $d^{cand}_l$) such that $d^{cand}_l = d^{pub} + l$, where $l = -10, \ldots , -1, 0, +1, \ldots , +10$, limiting the possible date references that are considered within 10 days before or after the tweet creation date, in addition to $d^{cand}_l = null$ (the tweet does not mention a date). 
\footnote{Although the temporal recognizer is trained with tweets from $\pm7$ days around the event date, we found that extending the candidate date range to $\pm10$ days for the temporal normalizer increased the performance of TweeTIME in the dev set.}
While our basic approach has the limitation, that it is only able to predict dates within $\pm 10$ days of the target date, we found that in practice the majority of date references on social media fall within this window.  Our approach is also able to score dates outside this range that are generated by traditional approaches to resolving time expressions, as described in Section \ref{sec:system_combination}.

The normalizer is similarly trained using the event database as distant supervision. The probability that a tweet mentions a candidate date is estimated using a log-linear model:
\begin{equation}
    P(d^{cand} | \textbf{w}, d^{pub}) 	\propto e^{\bm{\theta^n} \cdot \textbf{g}(\textbf{w}, d^{pub}, \textbf{t})}
\end{equation}
where $\bm{\theta^n}$ and $\textbf{g}$ are the parameter and feature vector respectively in the Temporal Normalizer. For every tweet and candidate date pair ($\textbf{w}$, $d^{cand}_l$), we extract the following set of features:
\newline

\vspace{-.13in}

\noindent {\bf{Temporal Tag Features}} that indicate whether the candidate date agrees with the temporal tags extracted by the Temporal Recognizer. Three cases can happen here: The recognizer can extract a tag that can not be derived from the candidate date; The recognizer can miss a tag derived from the candidate date; The recognizer can extract a tag that is derived from the candidate date. 
\newline 

\vspace{-.14in}

\noindent {\bf{Lexical Features}} that include two types of binary features from the tweet: 1) \textbf{Word Tag} features consist of conjunctions of words in the tweet and tags associated with the candidate date.  We remove URLs, stop words and punctuation; 2) \textbf{Word POS} features that are the same as above, but include conjunctions of POS tags, words and temporal tags derived from the candidate date.
\newline

\vspace{-.14in}

\noindent {\bf{Time Difference Features}} are numerical features that indicate the distance between the creation date and the candidate date.  They include difference of day ranges form -10 to 10 and the difference of week ranges from -2 to 2.

\section{Experiments}


In the following sub-sections we present experimental results on learning to resolve time expressions in Twitter using minimal supervision.  We start by describing our dataset, and proceed to present our results, including a large-scale evaluation on heuristically-labeled data and an evaluation comparing against human judgements.

\subsection{Data Collection}
\label{sec:event_extraction}

We collected  around $120$ million tweets posted in a one year window starting from April 2011 to May 2012. These tweets were automatically annotated with named entities, POS tags and TempEx dates \cite{ritter2011named}.


From this automatically-annotated corpus we extract the top $10,000$ events and their corresponding dates using the $G^2$ test, which measures the strength of association between an entity and date using the log-likelihood ratio between a model in which the entity is conditioned on the date and a model of independence \cite{ritter2012open}. Events extracted using this approach then simply consist of the highest-scoring entity-date pairs, for example [{\em Mercury, 5/9/2016}]. 

After automatically extracting the database of events, we next gather all tweets that mention an entity from the list that are also written within $\pm7$ days of the event.
These tweets and the dates of the known events serve as labeled examples that are likely to mention a known date. 

We also include a set of pseudo-negative examples, that are unlikely to refer to any event, by gathering a random sample of tweets that do not mention any of the top $10,000$ events and where TempEx does not extract any date.

\subsection{Large-Scale Heuristic Evaluation}
We first evaluate our tagging model, by testing how well it can predict the heuristically generated labels.  As noted in previous work on distant supervision \cite{Mintz_distantsupervision}, this type of evaluation usually under-estimates precision, however it provides us with a useful intrinsic measure of performance.

In order to provide even coverage of months in the training and test set, we divide the twitter corpus into 3 subsets based on the mod-5 week of each tweet's creation date. To train system we use tweets that are created in $1st$, $2nd$ or $3rd$ weeks. To tune parameters of the MiDaT model we used tweets from $5th$ weeks, and to evaluate the performance of the trained model we used tweets from $4th$ weeks.

\begin{table}[!ht]
\centering
\begin{tabular}{c |c| c| c} 
 \hline
   & \textbf{Precision} & \textbf{Recall} & \textbf{F-value} \\ 
 \hline\hline
 MultiT & 0.61 & 0.21 & 0.32 \\ 
 \hline
 MiDaT & 0.67 & 0.31 & 0.42 \\ 
 \hline
\end{tabular}
\caption{Performance comparison of MultiT and MiDaT at predicting heuristically generated tags on the dev set.}
\label{tab:PRF_mutir_dnmar}
\end{table}

The performance of the MiDaT model varies with the penalty and reward parameters. To find a (near) optimal setting of the values we performed a grid search on the dev set and found that a penalty of $-25$ and reward of 500 works best. A comparison of MultiT and MiDaT's performance at predicting heuristically generated labels is shown in Table \ref{tab:PRF_mutir_dnmar}.

The word level tags predicted by the temporal recognizer are used as the input to the temporal normalizer, which predicts the referenced date from each tweet. The overall system's performance at predicting event dates on the automatically generated test set, compared against SUTime, is shown in Table \ref{tab:comp_w_sutime_test_automated}.
    
\begin{table}[!ht]
\centering
\begin{tabular}{c|l | c|c|c}
 \hline
  & \textbf{System} & \textbf{Prec.} & \textbf{Recall} & \textbf{F-value} \\
 \hline\hline
 \multirow{2}{*}{dev set} &TweeTIME & 0.93 & 0.69 & 0.79 \\ 
 \hhline{~----}
   &SUTime & 0.89 & 0.64 & 0.75 \\
 \hline
 \multirow{2}{*}{test set} &TweeTIME & 0.97 & 0.94 & 0.96 \\ 
 \hhline{~----}
   &SUTime & 0.85 & 0.75 & 0.80 \\  
 \hline
\end{tabular}
\caption{Performance comparison of TweeTIME and SUTime at predicting heuristically labeled normalized dates.}
\label{tab:comp_w_sutime_test_automated}
\end{table}

\begin{table*}[!ht]
\centering
\begin{tabular}{p{6.5 in}} 
 \hline
   \textbf{Tweets and their corresponding word tags ($\text{word}^{tag}$)}\\
 \hline\hline
 $\text{Im}^{NA}\  \text{hella}^{NA}\  \text{excited}^{future}\  \text{for}^{NA}\  \text{tomorrow}^{future}$\\
 \hline
 $\text{Kick}^{NA}\  \text{off}^{NA}\  \text{the}^{NA}\  \text{New}^{future}\  \text{Year}^{future}\  \text{Right}^{NA}\  \text{@}^{NA}\  \text{\#ClubLacura}^{NA}\  \text{\#FRIDAY}^{fri}\ \text{!}^{NA}\  $  \\
 $\text{HOSTED}^{NA}\ \text{BY}^{NA}\  \text{[[}^{NA}\  \text{DC}^{NA}\  \text{Young}^{NA}\  \text{Fly}^{NA}\  \text{]]}^{NA}$\\
 \hline
 @$\text{OxfordTownHall}^{NA}\ \text{Thks}^{NA}\ \text{for}^{NA}\ \text{a}^{NA}\ \text{top}^{NA}\ \text{night}^{NA}\ \text{at}^{NA}\ \text{our}^{NA}\ \text{Christmas}^{dec}\ \text{party}^{NA}\ \text{on}^{NA}\ \text{Fri!}^{fri}$ \\ 
 $\text{Compliments}^{NA}\ \text{to}^{NA}\ \text{chef!}^{NA}\ \text{(Rose}^{NA}\ \text{melon}^{NA}\ cantaloupe^{NA}\ :)^{NA}$\\
 \hline
 $\text{Im}^{NA}\  \text{proud}^{NA}\ \text{to}^{NA}\ \text{say}^{NA}\ \text{that}^{NA}\ \text{I}^{NA}\  \text{breathed}^{past}\ \text{the}^{NA}\ \text{same}^{NA}\ \text{air}^{NA}\ \text{as}^{NA}\ \text{Harry}^{NA}\ \text{on}^{NA}\ \text{March}^{mar}\  $ \\
 $\text{21,}^{21}\ \text{2015.}^{NA}\  \text{\#KCA}^{NA}\ \text{\#Vote1DUK}^{NA}\  $\\
\hline
$\text{C'mon}^{present}\ \text{let's}^{present}\  \text{jack}^{NA}\ \text{Tonight}^{present}\ \text{will}^{NA}\ \text{be}^{present}\  \text{a}^{NA}\ \text{night}^{NA}\ \text{to}^{NA}\ \text{remember.}^{NA}$\\
\hline
 
\end{tabular}
\caption{Example MiDaT tagging output on the test set.}
\label{tab:wor_tag}
\end{table*}

\begin{table}[!ht]
\centering
\begin{tabular}{p{1 in} | c |c| c} 
 \hline
   & \textbf{Precision} & \textbf{Recall} & \textbf{F-value} \\
 \hline\hline
 TweeTIME & 0.61 & 0.81 & 0.70 \\ 
  \hline
 - Day Diff. & 0.46 & 0.72 & 0.56 \\
\hline
  - Lexical\&POS & 0.48 & 0.80 & 0.60 \\ 
 \hline
  - Week Diff. & 0.49 & 0.85 & 0.62 \\ 
 \hline
 - Lexical & 0.50 & 0.88 & 0.64 \\ 
 \hline
  - Temporal Tag & 0.57 & 0.83 & 0.68 \\ 
 \hline
\end{tabular}
\caption{Feature ablation of the Temporal Resolver by removing each individual feature group from the full set.}
\label{tab:PRF_feature_date_resolver}
\end{table}

\subsection{Evaluation Against Human Judgements}

In addition to automatically evaluating our tagger on a large corpus of heuristically-labeled tweets, we also evaluate the performance of our tagging and date-resolution models on a random sample of tweets taken from a much later time period, that were manually annotated by the authors.

\subsubsection{Word-Level Tags}
To evaluate the performance of the MiDaT-tagger we randomly selected 50 tweets and labeled each word with its corresponding tag.  Against this hand annotated test set, MiDaT achieves Precision=0.54, Recall=0.45 and  F-value=0.49. A few examples of word-level tags predicted by MiDaT are shown in Table \ref{tab:wor_tag}.  We found that because the tags are learned as latent variables inferred by our model, they sometimes don't line up exactly with our intuitions but still provide useful predictions, for example in Table \ref{tab:wor_tag}, {\em Christmas} is labeled with the tag MOY=$dec$.

\begin{table}[!ht]
\centering
\begin{tabular}{p{.2in}|l| c|c|c}
 \hline
  & \textbf{System} & \textbf{Prec.} & \textbf{Recall} & \textbf{F-value} \\
 \hline\hline
  \multirow{6}{*}{\parbox{.2in}{dev set}}& TweeTIME & \textbf{0.61} & 0.81 & \textbf{0.70} \\ 
   \hhline{~----}
 & TweeTIME+SU & \textbf{0.67} & \textbf{0.83} & \textbf{0.74} \\ 
 \hhline{~----}
 & SUTime & 0.51 & \textbf{0.86} & 0.64 \\  
 \hhline{~----}
 & TempEx & 0.58 & 0.64 & 0.61 \\  
 \hhline{~----}
 & HeidelTime & 0.57 & 0.63 & 0.60 \\
 \hhline{~----}
 & UWTime & 0.49 & 0.57 & 0.53 \\

 \hline
 \multirow{6}{*}{\parbox{.2in}{test set}}& TweeTIME & \textbf{0.58} & \textbf{0.70} & \textbf{0.63} \\  
 \hhline{~----}
 & TweeTIME+SU & \textbf{0.62} & \textbf{0.76} & \textbf{0.68} \\
 \hhline{~----}
 & SUTime & 0.54 & 0.64 & 0.58 \\  
 \hhline{~----}
 & TempEx & 0.56 & 0.58 & 0.57 \\  
 \hhline{~----}
 & HeidelTime & 0.43 & 0.52 & 0.47 \\  
 \hhline{~----}
 & UWTime & 0.39 & 0.50 & 0.44\\
 
 \hline
\end{tabular}
\caption{Performance comparison of TweeTIME against state-of-the-art temporal taggers. TweeTIME+SU uses our proposed approach to system combination, re-scoring output from SUTime using extracted features and learned parameters from TweeTIME.}
\label{tab:comp_w_prev_sys}
\end{table}

\subsubsection{End-to-end Date Resolution}
To evaluate the final performance of our system and compare against existing state-of-the art time resolvers, we randomly sampled 250 tweets from 2014-2016 and manually annotated them with normalized dates; note that this is a separate date range from our weakly-labeled training data which is taken from 2011-2012. We use 50 tweets as a development set and the remaining 200 as a final test set. We experimented with different feature sets on the development data.  Feature ablation experiments are presented in Table \ref{tab:PRF_feature_date_resolver}.

\begin{figure}[!ht]
\centering{
    \includegraphics[width=0.4 \textwidth]{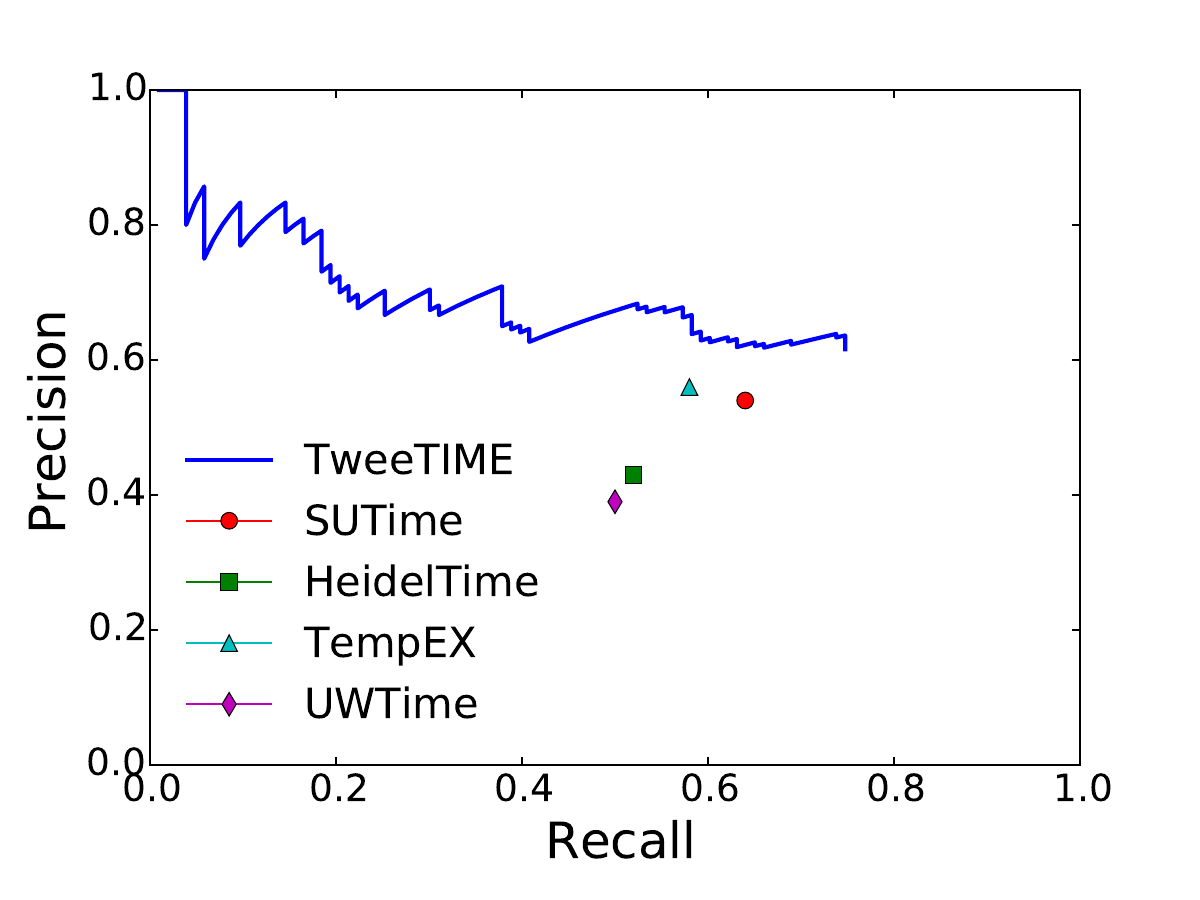}
    \caption{Precision and recall at resolving time expressions compared against human judgements. TweeTIME achieves higher precision at comparable recall than other state-of-the-art systems. }
    \label{pp_plot}
}
\end{figure}

The final performance of our system, compared against a range of state-of-the-art time resolvers is presented in Table \ref{tab:comp_w_prev_sys}.  We see that TweeTIME outperforms SUTime, Tempex, HeidelTime (using its COLLOQUIAL mode, which is designed for SMS text) and UWTime. Brief descriptions of each system can be found in Section \ref{sec:relatedwork}.

\subsubsection{System Combination with SUTime}
\label{sec:system_combination}

As our basic TweeTIME system is designed to predict dates within $\pm 10$ days of the creation date, it fails when a tweet refers to a date outside this range. To overcome this limitation we append the date predicted by SUTime in the list of candidate days.  We then re-rank SUTime's predictions using our log-linear model, and include its output as a predicted date if the confidence of our normalizer is sufficiently high.

\subsubsection{Error Analysis}

\begin{figure*}[htbp]
    \centering
    \includegraphics[width=\textwidth]{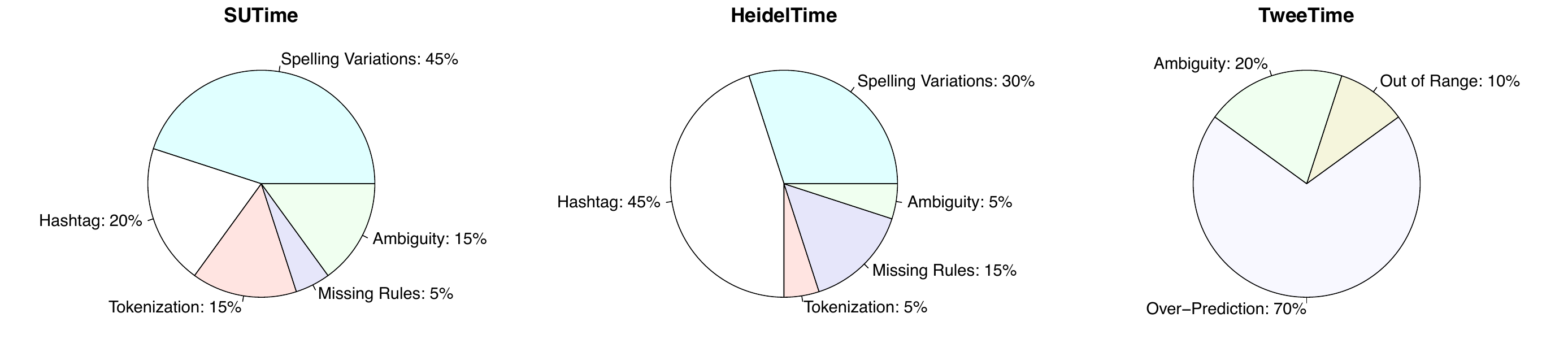}
    \caption{Error analyses for different temporal resolvers}
    \label{fig:error_analysis}

\end{figure*}

We manually examined the system outputs and found 7 typical categories of errors (see examples in Table \ref{tab:error_examples_alter}):

\noindent\textbf{Spelling Variation:} Twitter users are very creative in their use of spelling and abbreviations. For example, a large number of variations of the word {\em tomorrow} can be found in tweets, including {\em 2morrow}, {\em 2mrw}, {\em tmrw}, {\em 2mrow} and so on. Previous temporal resolvers often fail in these cases, while TweeTIME significantly reduces such errors.

\noindent\textbf{Ambiguity:} In many cases, temporal words like {\em Friday} in the tweet {\em Is it Friday yet?} may not refer to any specific event or date, but are often predicted incorrectly. Also included in this category are cases where the future and past are confused.  For example, predicting the past Friday, when it is actually the coming Friday.

\noindent\textbf{Missing Rule:} Cases where specific temporal keywords, such as {\em April Fools}, are not covered by the rule-based systems. 

\noindent\textbf{Tokenization:} Traditional systems tend to be very sensitive to incorrect tokenization and have trouble to handle expressions such as {\em 9th-december}, {\em May   9,2015} or {\em Jan1}. For the following Tweet:

\begin{quotation}
\noindent JUST IN Delhi high court asks state government  to  submit  data  on  changes  in  pollution level since \#OddEven rule came into effect on {\em Jan1}
\end{quotation}
TweeTIME is able to correctly extract {\em 01/01/2016}, whereas HeidelTime, SUTime, TempEX and UWTime all failed to extract any dates.

\noindent\textbf{Hashtag:} Hashtags can carry temporal information, for example, \#September11. Only our system that is adapted to social media can resolve these cases.

\noindent\textbf{Out of Range:\hspace{-0.0491 in}} TweeTIME only predicts dates within 10 days before or after the tweet. Time expressions referring to dates outside this range will not be predicted correctly. System combination with SUTime (Section \ref{sec:system_combination}) only partially addressed this problem. 


\noindent\textbf{Over-Prediction:} \hspace{-0.01in}Unlike rule-based systems, TweeTIME has a tendency to over-predict when there is no explicit time expression in the tweets, possibly because of the presence of present tense verbs. Such mistakes could also happen in some past tense verbs.

Because TweeTIME resolves time expressions using a very different approach compared to traditional methods, its distribution of errors is quite distinct, as illustrated in Figure \ref{fig:error_analysis}.

\begin{table*}[bhpt]
\centering
\small
\begin{tabular}{|p{1in}|p{2.7 in}|p{0.7in}|p{1.4in}|}

\hline
  {\centering Error Category} & {\centering Tweet} & {\centering Gold Date} & {\centering Predicted Date}\\ 
 \hline\hline
 \centering \multirow{2}* {\textbf{Spelling}} & {\centering \multirow{2}* {I cant believe \textit{tmrw} is \textit{fri}..the week flys by}} & {\centering \multirow{2}* {2015-03-06}} & None (SUTime, HeidelTime)\\
\hline
  
  \centering \multirow{2}*{\textbf{Ambiguity}} & {\centering \multirow{2}* { RT @Iyaimkatie: Is it Friday yet?????}} & {\centering \multirow{2}* {None}} & 2015-12-04 (TweeTime, SUTime, HeidelTime) \\
\hline
\centering \multirow{2}*{\textbf{Missing Rule}}  & \#49ers \#sanfrancisco 49ers fans should be oh so wary of \textit{April Fools} pranks & {\centering \multirow{2}* {2015-04-01}} & {\centering \multirow{2}* {None (HeidelTime)}} \\
\hline
  \centering \multirow{2}*{\textbf{Tokenization}} & 100000 - still waiting for that reply from \textit{9th-december} lmao. you're pretty funny and chill & {\centering \multirow{2}* {2015-12-09}} & None (SUTime, HeidelTime)\\
\hline
   \centering \multirow{2}*{\textbf{Hashtag}} & RT @arianatotally: Who listening to the \textit{\#SATURDAY} \#Night w/ @AlexAngelo?I'm loving it. & {\centering \multirow{2}* {2015-04-11}} & None (SUTime, HeidelTime)\\
   
\hline

   \centering \multirow{3}*{\textbf{Out of Range}}  &  RT @460km: In memory of Constable Christine Diotte @rcmpgrcpolice EOW: \textit{March 12, 2002} \#HeroesInLife \#HerosEnVie  & \centering \multirow{3}* {2002-03-12} & {\centering \multirow{3}* {2015-03-12 (TweeTime)}}\\
\hline
\centering \multirow{2}* {\textbf{Over-Prediction}} & RT @tinatbh: January 2015: this will be my year December 2015: maybe not. & {\centering \multirow{2}* {None}} & {\centering \multirow{2}* {2015-12-08 (TweeTime)}} \\
\hline
\end{tabular}
\caption{Representative Examples of System (SUTime, HeidelTime, TweeTIME) Errors}
\label{tab:error_examples_alter}
\end{table*}

\section{Related Work}
\label{sec:relatedwork}

\noindent {\bf{Temporal Resolvers}} primarily utilize either rule-based or probabilistic approaches. Notable rule-based systems such as TempEx \cite{mani2000robust}, SUTime \cite{chang2012sutime} and HeidelTime \cite{strotgen2013multilingual} provide particularly competitive performance compared to the state-of-the-art machine learning methods. Probabilistic approaches use supervised classifiers trained on in-domain annotated data \cite{kolomiyets2010kul,bethard:2013:SemEval-2013,Filannino:2013:SemEval-2013} or hybrid with hand-engineered rules \cite{uzzaman2010trips,lee2014context}. UWTime \cite{lee2014context} is one of the most recent and competitive systems and uses Combinatory Categorial Grammar (CCG). 

Although the recent research challenge TempEval \cite{uzzaman2013tempeval,bethard2016tempeval} offers an evaluation in the clinical domain besides newswire, most participants used the provided annotated corpus to train supervised models in addition to employing hand-coded rules. Previous work on adapting temporal taggers primarily focus on scaling up to more languages. HeidelTime was extended to multilingual \cite{strotgen-gertz:2015:EMNLP}, colloquial (SMS) and scientific texts \cite{strotgen2012temporal} using dictionaries and additional in-domain annotated data. One existing work used distant supervision \cite{angeli2012parsing,angeli2013language}, but for normalization only, assuming gold time mentions as input. They used an EM-style bootstrapping approach and a CKY parser. 

\noindent {\bf{Distant Supervision}} has recently become popular in natural language processing. Much of the work has focused on the task of relation extraction \cite{craven99,bunescu07,mintz09,riedel10,hoffmann11,nguyen11,surdeanu12,xu13,ritter13,angeli2014combining}. Recent work also shows exciting results on extracting named entities \cite{ritter2011named,plank-EtAl:2014:Coling}, emotions \cite{purver2012experimenting}, sentiment \cite{marchetti2012learning}, as well as finding evidence in medical publications \cite{wallaceextracting}. Our work is closely related to the joint word-sentence model that exploits multiple-instance learning for paraphrase identification \cite{Xu-EtAl-2014:TACL} in Twitter. 

\section{Conclusions}
\vspace{-.088 in}
In this paper, we showed how to learn time resolvers from large amounts of unlabeled text, using a database of known events as distant supervision.  We presented a method for learning a word-level temporal tagging models from tweets that are heuristically labeled with only sentence-level labels.  This approach was further extended to account for the case of missing tags, or temporal properties that are not explicitly mentioned in the text of a tweet.  These temporal tags were then combined with a variety of other features in a novel date-resolver that predicts normalized dates referenced in a Tweet.  By learning from large quantities of in-domain data, we were able to achieve 0.68 F1 score on the end-to-end time normalization task for social media data, significantly outperforming SUTime, TempEx, HeidelTime and UWTime on this challenging dataset for time normalization.

\section*{Acknowledgments}
We would like to thank the anonymous reviewers for helpful feedback on a previous draft.  This material is based upon work supported by the National Science Foundation under Grant No. IIS-1464128.  Alan Ritter is supported by the Office of the Director of National Intelligence (ODNI) and the Intelligence Advanced Research Projects Activity (IARPA) via the Air Force Research Laboratory (AFRL) contract number FA8750-16-C-0114. The U.S. Government is authorized to reproduce and distribute reprints for Governmental purposes notwithstanding any copyright annotation thereon. Disclaimer: The views and conclusions contained herein are those of the authors and should not be interpreted as necessarily representing the official policies or endorsements, either expressed or implied, of ODNI, IARPA, AFRL, or the U.S. Government.


\bibliography{emnlp2016}

\begin{thebibliography}{}

\bibitem[\protect\citename{Alonso \bgroup et al.\egroup }2007]{alonso2007value}
Omar Alonso, Michael Gertz, and Ricardo Baeza-Yates.
\newblock 2007.
\newblock On the value of temporal information in information retrieval.
\newblock In {\em ACM SIGIR Forum}, volume~41, pages 35--41. ACM.

\bibitem[\protect\citename{Angeli and Uszkoreit}2013]{angeli2013language}
Gabor Angeli and Jakob Uszkoreit.
\newblock 2013.
\newblock Language-independent discriminative parsing of temporal expressions.
\newblock In {\em Proceedings of the 51st Annual Meeting of the Association for
  Computational Linguistics (ACL)}.

\bibitem[\protect\citename{Angeli \bgroup et al.\egroup
  }2012]{angeli2012parsing}
Gabor Angeli, Christopher~D Manning, and Daniel Jurafsky.
\newblock 2012.
\newblock Parsing time: Learning to interpret time expressions.
\newblock In {\em Proceedings of the 2012 Conference of the North American
  Chapter of the Association for Computational Linguistics: Human Language
  Technologies (NAACL)}.

\bibitem[\protect\citename{Angeli \bgroup et al.\egroup
  }2014]{angeli2014combining}
Gabor Angeli, Julie Tibshirani, Jean Wu, and Christopher~D Manning.
\newblock 2014.
\newblock Combining distant and partial supervision for relation extraction.
\newblock In {\em Proceedings of the 2014 Conference on Empirical Methods in
  Natural Language Processing (EMNLP)}.

\bibitem[\protect\citename{Bethard and Savova}2016]{bethard2016tempeval}
Steven Bethard and Guergana Savova.
\newblock 2016.
\newblock {SemEval-2016 Task 12: Clinical TempEval}.
\newblock In {\em Proceedings of the 10th International Workshop on Semantic
  Evaluation (SemEval)}.

\bibitem[\protect\citename{Bethard}2013a]{bethard:2013:SemEval-2013}
Steven Bethard.
\newblock 2013a.
\newblock {ClearTK-TimeML: A} minimalist approach to {TempEval} 2013.
\newblock In {\em Proceedings of the Seventh International Workshop on Semantic
  Evaluation (SemEval)}.

\bibitem[\protect\citename{Bethard}2013b]{bethard:2013:EMNLP}
Steven Bethard.
\newblock 2013b.
\newblock A synchronous context free grammar for time normalization.
\newblock In {\em Proceedings of the 2013 Conference on Empirical Methods in
  Natural Language Processing (EMNLP)}.

\bibitem[\protect\citename{Bunescu and Mooney}2007]{bunescu07}
Razvan~C. Bunescu and Raymond~J. Mooney.
\newblock 2007.
\newblock Learning to extract relations from the {W}eb using minimal
  supervision.
\newblock In {\em Proceedings of the 45th Annual Meeting of the Association for
  Computational Linguistics (ACL)}.

\bibitem[\protect\citename{Chambers}2013]{chambers2013tempeval}
Nathanael Chambers.
\newblock 2013.
\newblock {NavyTime: E}vent and time ordering from raw text.
\newblock In {\em Proceedings of the 7th International Workshop on Semantic
  Evaluation (SemEval)}.

\bibitem[\protect\citename{Chang and Manning}2012]{chang2012sutime}
Angel~X Chang and Christopher~D Manning.
\newblock 2012.
\newblock {SUTime: A} library for recognizing and normalizing time expressions.
\newblock In {\em Proceedings of the 8th International Conference on Language
  Resources and Evaluation (LREC)}.

\bibitem[\protect\citename{Chang \bgroup et al.\egroup }2016]{changexpectation}
Ching-Yun Chang, Zhiyang Teng, and Yue Zhang.
\newblock 2016.
\newblock Expectation-regulated neural model for event mention extraction.
\newblock {\em Proccedings of the 2016 Conference of the North American Chapter
  of the Association for Computational Linguistics: Technologies (NAACL)}.

\bibitem[\protect\citename{Craven and Kumlien}1999]{craven99}
Mark Craven and Johan Kumlien.
\newblock 1999.
\newblock Constructing biological knowledge bases by extracting information
  from text sources.
\newblock In {\em Proceedings of the Seventh International Conference on
  Intelligent Systems for Molecular Biology (ISMB)}.

\bibitem[\protect\citename{Derczynski and
  Gaizauskas}2013]{derczynski2013temporal}
Leon Derczynski and Robert~J Gaizauskas.
\newblock 2013.
\newblock Temporal signals help label temporal relations.
\newblock In {\em Proceedings of the 51st Annual Meeting of the Association for
  Computational Linguistics (ACL)}.

\bibitem[\protect\citename{Dietterich \bgroup et al.\egroup
  }1997]{dietterich1997solving}
Thomas~G Dietterich, Richard~H Lathrop, and Tom{\'a}s Lozano-P{\'e}rez.
\newblock 1997.
\newblock Solving the multiple instance problem with axis-parallel rectangles.
\newblock {\em Artificial intelligence}, 89(1).

\bibitem[\protect\citename{Do \bgroup et al.\egroup }2012]{Do2012}
Quang~Xuan Do, Wei Lu, and Dan Roth.
\newblock 2012.
\newblock Joint inference for event timeline construction.
\newblock In {\em Proceedings of the 2012 Joint Conference on Empirical Methods
  in Natural Language Processing and Computational Natural Language Learning
  (EMNLP)}.

\bibitem[\protect\citename{Filannino \bgroup et al.\egroup
  }2013]{Filannino:2013:SemEval-2013}
Michele Filannino, Gavin Brown, and Goran Nenadic.
\newblock 2013.
\newblock {ManTIME: T}emporal expression identification and normalization in
  the {TempEval}-3 challenge.
\newblock In {\em Proceedings of the Seventh International Workshop on Semantic
  Evaluation (SemEval)}.

\bibitem[\protect\citename{Hoffmann \bgroup et al.\egroup
  }2011a]{Hoffmann_multir}
Raphael Hoffmann, Congle Zhang, Xiao Ling, Luke Zettlemoyer, and Daniel~S.
  Weld.
\newblock 2011a.
\newblock Knowledge-based weak supervision for information extraction of
  overlapping relations.
\newblock In {\em The 49th Annual Meeting of the Association for Computational
  Linguistics: Human Language Technologies (ACL)}.

\bibitem[\protect\citename{Hoffmann \bgroup et al.\egroup }2011b]{hoffmann11}
Raphael Hoffmann, Congle Zhang, Xiao Ling, Luke~S. Zettlemoyer, and Daniel~S.
  Weld.
\newblock 2011b.
\newblock Knowledge-based weak supervision for information extraction of
  overlapping relations.
\newblock In {\em Proceedings of the 49th Annual Meeting of the Association for
  Computational Linguistics (ACL)}.

\bibitem[\protect\citename{Ji \bgroup et al.\egroup }2011]{ji2011overview}
Heng Ji, Ralph Grishman, Hoa~Trang Dang, Kira Griffitt, and Joe Ellis.
\newblock 2011.
\newblock Overview of the tac 2011 knowledge base population track.
\newblock In {\em Proceedings of the Fourth Text Analysis Conference (TAC)}.

\bibitem[\protect\citename{Kanhabua \bgroup et al.\egroup
  }2012]{kanhabua2012supporting}
Nattiya Kanhabua, Sara Romano, Avar{\'e} Stewart, and Wolfgang Nejdl.
\newblock 2012.
\newblock Supporting temporal analytics for health-related events in
  microblogs.
\newblock In {\em Proceedings of the 21st ACM International Conference on
  Information and Knowledge Management (CIKM)}.

\bibitem[\protect\citename{Kolomiyets and Moens}2010]{kolomiyets2010kul}
Oleksandr Kolomiyets and Marie-Francine Moens.
\newblock 2010.
\newblock {KUL}: {R}ecognition and normalization of temporal expressions.
\newblock In {\em Proceedings of the 5th International Workshop on Semantic
  Evaluation (SemEval)}.

\bibitem[\protect\citename{Lee \bgroup et al.\egroup }2014]{lee2014context}
Kenton Lee, Yoav Artzi, Jesse Dodge, and Luke Zettlemoyer.
\newblock 2014.
\newblock Context-dependent semantic parsing for time expressions.
\newblock In {\em Proceedings of 52nd Annual Meeting of the Association for
  Computational Linguistics (ACL)}.

\bibitem[\protect\citename{Ling and Weld}2010]{ling2010temporal}
Xiao Ling and Daniel~S Weld.
\newblock 2010.
\newblock Temporal information extraction.
\newblock In {\em Proceedings of the 24th AAAI Conference on Artificial
  Intelligence (AAAI)}.

\bibitem[\protect\citename{Mani and Wilson}2000]{mani2000robust}
Inderjeet Mani and George Wilson.
\newblock 2000.
\newblock Robust temporal processing of news.
\newblock In {\em Proceedings of the 38th Annual Meeting on Association for
  Computational Linguistics (ACL)}.

\bibitem[\protect\citename{Marchetti-Bowick and
  Chambers}2012]{marchetti2012learning}
Micol Marchetti-Bowick and Nathanael Chambers.
\newblock 2012.
\newblock Learning for microblogs with distant supervision: {P}olitical
  forecasting with {T}witter.
\newblock In {\em Proceedings of the 13th Conference of the European Chapter of
  the Association for Computational Linguistics (EACL)}.

\bibitem[\protect\citename{Mintz \bgroup et al.\egroup
  }2009a]{Mintz_distantsupervision}
Mike Mintz, Steven Bills, Rion Snow, and Dan Jurafsky.
\newblock 2009a.
\newblock Distant supervision for relation extraction without labeled data.
\newblock In {\em Proceedings of the Joint Conference of the Association of
  Computational Linguistics and the International Joint Conference on Natural
  Language Processing (ACL-IJCNLP)}.

\bibitem[\protect\citename{Mintz \bgroup et al.\egroup }2009b]{mintz09}
Mike Mintz, Steven Bills, Rion Snow, and Daniel Jurafsky.
\newblock 2009b.
\newblock Distant supervision for relation extraction without labeled data.
\newblock In {\em Proceedigns of the 47th Annual Meeting of the Association for
  Computational Linguistics and the 4th International Joint Conference on
  Natural Language Processing (ACL)}.

\bibitem[\protect\citename{Nguyen and Moschitti}2011]{nguyen11}
Truc-Vien~T. Nguyen and Alessandro Moschitti.
\newblock 2011.
\newblock End-to-end relation extraction using distant supervision from
  external semantic repositories.
\newblock In {\em Proceedings of the 49th Annual Meeting of the Association for
  Computational Linguistics (ACL)}.

\bibitem[\protect\citename{Plank \bgroup et al.\egroup
  }2014]{plank-EtAl:2014:Coling}
Barbara Plank, Dirk Hovy, Ryan McDonald, and Anders S{\o}gaard.
\newblock 2014.
\newblock Adapting taggers to twitter with not-so-distant supervision.
\newblock pages 1783--1792.

\bibitem[\protect\citename{Purver and Battersby}2012]{purver2012experimenting}
Matthew Purver and Stuart Battersby.
\newblock 2012.
\newblock Experimenting with distant supervision for emotion classification.
\newblock In {\em Proceedings of the 13th Conference of the European Chapter of
  the Association for Computational Linguistics (EACL)}.

\bibitem[\protect\citename{Riedel \bgroup et al.\egroup }2010]{riedel10}
Sebastian Riedel, Limin Yao, and Andrew McCallum.
\newblock 2010.
\newblock Modeling relations and their mentions without labeled text.
\newblock In {\em Proceedigns of the European Conference on Machine Learning
  and Principles and Practice of Knowledge Discovery in Databases (ECML-PKDD)}.

\bibitem[\protect\citename{Ritter \bgroup et al.\egroup }2011]{ritter2011named}
Alan Ritter, Mausam, Sam Clark, and Oren Etzioni.
\newblock 2011.
\newblock Named entity recognition in {T}weets: {A}n experimental study.
\newblock In {\em Proceedings of the Conference on Empirical Methods in Natural
  Language Processing (EMNLP)}.

\bibitem[\protect\citename{Ritter \bgroup et al.\egroup }2012]{ritter2012open}
Alan Ritter, Mausam, Oren Etzioni, and Sam Clark.
\newblock 2012.
\newblock Open domain event extraction from twitter.
\newblock In {\em Proceedings of the 18th ACM SIGKDD international conference
  on Knowledge discovery and data mining (KDD)}.

\bibitem[\protect\citename{Ritter \bgroup et al.\egroup }2013]{ritter13}
Alan Ritter, Luke Zettlemoyer, Mausam, and Oren Etzioni.
\newblock 2013.
\newblock Modeling missing data in distant supervision for information
  extraction.
\newblock {\em Transactions of the Association for Computational Linguistics
  (TACL)}, 1:367--378.

\bibitem[\protect\citename{Ritter \bgroup et al.\egroup
  }2015]{ritter2015weakly}
Alan Ritter, Evan Wright, William Casey, and Tom Mitchell.
\newblock 2015.
\newblock Weakly supervised extraction of computer security events from
  {T}witter.
\newblock In {\em Proceedings of the 24th International Conference on World
  Wide Web (WWW)}.

\bibitem[\protect\citename{Schwartz \bgroup et al.\egroup
  }2015]{schwartz2015extracting}
H~Andrew Schwartz, Greg Park, Maarten Sap, Evan Weingarten, Johannes
  Eichstaedt, Margaret Kern, Jonah Berger, Martin Seligman, and Lyle Ungar.
\newblock 2015.
\newblock Extracting human temporal orientation in {F}acebook language.
\newblock In {\em Proceedings of the 2015 Conference of the North American
  Chapter of the Association for Computational Linguistics: Human Language
  Technologies (NAACL)}.

\bibitem[\protect\citename{Str{\"o}tgen and Gertz}2012]{strotgen2012temporal}
Jannik Str{\"o}tgen and Michael Gertz.
\newblock 2012.
\newblock Temporal tagging on different domains: {C}hallenges, strategies, and
  gold standards.
\newblock In {\em Proceedings of the 8th International Conference on Language
  Resources and Evaluation (LREC)}.

\bibitem[\protect\citename{Str{\"o}tgen and
  Gertz}2013]{strotgen2013multilingual}
Jannik Str{\"o}tgen and Michael Gertz.
\newblock 2013.
\newblock Multilingual and cross-domain temporal tagging.
\newblock {\em Language Resources and Evaluation}, 47(2):269--298.

\bibitem[\protect\citename{Str\"{o}tgen and
  Gertz}2015]{strotgen-gertz:2015:EMNLP}
Jannik Str\"{o}tgen and Michael Gertz.
\newblock 2015.
\newblock A baseline temporal tagger for all languages.
\newblock In {\em Proceedings of the 2015 Conference on Empirical Methods in
  Natural Language Processing (EMNLP)}.

\bibitem[\protect\citename{Surdeanu \bgroup et al.\egroup }2012]{surdeanu12}
Mihai Surdeanu, Julie Tibshirani, Ramesh Nallapati, and Christopher~D. Manning.
\newblock 2012.
\newblock Multi-instance multi-label learning for relation extraction.
\newblock In {\em Proceedings of the 50th Annual Meeting of the Association for
  Computational Linguistics (ACL)}.

\bibitem[\protect\citename{UzZaman and Allen}2010]{uzzaman2010trips}
Naushad UzZaman and James~F Allen.
\newblock 2010.
\newblock {TRIPS} and {TRIOS} system for {T}empeval-2: {E}xtracting temporal
  information from text.
\newblock In {\em Proceedings of the 5th International Workshop on Semantic
  Evaluation (SemEval)}.

\bibitem[\protect\citename{UzZaman \bgroup et al.\egroup
  }2013]{uzzaman2013tempeval}
Naushad UzZaman, Hector Llorens, James Allen, Leon Derczynski, Marc Verhagen,
  and James Pustejovsky.
\newblock 2013.
\newblock {SemEval-2013 Task 1: TEMPEVAL-3: E}valuating time expressions,
  events, and temporal relations.
\newblock In {\em Proceedings of the 7th International Workshop on Semantic
  Evaluation (SemEval)}.

\bibitem[\protect\citename{Wallace \bgroup et al.\egroup
  }2016]{wallaceextracting}
Byron~C Wallace, Jo{\"e}l Kuiper, Aakash Sharma, Mingxi~Brian Zhu, and Iain~J
  Marshall.
\newblock 2016.
\newblock Extracting {PICO} sentences from clinical trial reports using
  supervised distant supervision.
\newblock {\em Journal of Machine Learning Research (JMLR)}.

\bibitem[\protect\citename{Xu \bgroup et al.\egroup }2013]{xu13}
Wei Xu, Raphael Hoffmann, Zhao Le, and Ralph Grishman.
\newblock 2013.
\newblock Filling knowledge base gaps for distant supervision of relation
  extraction.
\newblock In {\em Proceedings of the 51th Annual Meeting of the Association for
  Computational Linguistics (ACL)}.

\bibitem[\protect\citename{Xu \bgroup et al.\egroup }2014]{Xu-EtAl-2014:TACL}
Wei Xu, Alan Ritter, Chris Callison-Burch, William~B. Dolan, and Yangfeng Ji.
\newblock 2014.
\newblock Extracting lexically divergent paraphrases from {Twitter}.
\newblock {\em Transactions of the Association for Computational Linguistics
  (TACL)}, 2(1).

\end{thebibliography}
\bibliographystyle{emnlp2016}

\end{document}